\documentclass[conference]{IEEEtran}
\usepackage[boxruled]{algorithm2e}
\usepackage{cite}
\usepackage{graphicx}
\usepackage{psfrag}
\usepackage{subfigure}
\usepackage{url}
\usepackage{amsmath}
\usepackage{array}
\usepackage{amssymb}
\usepackage{amsfonts}
\usepackage{graphicx}
\usepackage{epstopdf}

\newtheorem{definition}{Definition}

\title{Bounds on the Achievable Rate for the Fading Relay Channel with Finite Input Constellations}
\begin{document}

\author{
\authorblockN{Vijayvaradharaj T Muralidharan}
\authorblockA{Dept. of ECE, Indian Institute of Science \\
Bangalore 560012, India\\
Email: tmvijay@ece.iisc.ernet.in
}
\and
\authorblockN{B. Sundar Rajan}
\authorblockA{Dept. of ECE, Indian Institute of Science, \\Bangalore 560012, India\\
Email: bsrajan@ece.iisc.ernet.in
}
}

\maketitle
\thispagestyle{empty}	
\begin{abstract}
We consider the wireless Rayleigh fading relay channel with finite complex input constellations. Assuming global knowledge of the channel state information and perfect synchronization,  upper and lower bounds on the achievable rate, for the full-duplex relay, as well as the more practical half-duplex relay (in which the relay cannot transmit and receive simultaneously), are studied. Assuming the power constraint at the source node and the relay node to be equal, the gain in rate offered by the use of relay over the direct transmission (without the relay) is investigated.  It is shown that for the case of finite complex input constellations, the relay gain attains the  maximum at a particular SNR and at higher SNRs the relay gain tends to become zero. Since practical schemes always use finite complex input constellation, the above result means that the relay offers maximum advantage over the direct transmission when we operate at a particular SNR and offers no advantage at very high SNRs. This is contrary to the results already known for the relay channel with Gaussian input alphabet.
\end{abstract}
\section{INTRODUCTION}
\label{sec1}
The study of the wireless relay channel has attracted a lot of attention in recent times. Achievable rates  and the upper bounds on the capacity for the full duplex (FD) discrete memoryless relay channel (DMC) were studied in \cite{CoEl}. Motivated by the practical constraints involved in the construction of relays which can transmit and receive simultaneously, the bounds on the capacity for the so-called cheap relay networks were derived \cite{KhSaAa1}. We will refer to such a relay channel as the half-duplex (HD) relay channel in this paper. These results were extended to the Gaussian relay channel with continuous input and output alphabets \cite{KhSaAa2}.

For a single-user Gaussian channel, if the input is constrained to come from a finite complex constellation with the points used with uniform distribution, the mutual information between the input and the output of the channel is called the Constellation Constrained Capacity (CCC) in \cite{Big}. Similar notion of CCC was studied for the Gaussian-MAC channel in \cite{HaR} and for the Gaussian Interference Channel in \cite{AbR} with finite complex input constellations. In this paper we study bounds (both upper and lower bounds) on the achievable rate of the relay channel with finite complex input constellations. 

The lower bound on the achievable rate of the relay channel is defined as be the achievable rate of the decode and forward scheme \cite{CoEl}, when the input symbols take values from a finite constellation. Similarly, the cut-set bound given by the max-flow min cut theorem \cite{Cover}, when the input symbols take values from a finite constellation, is referred to as the upper bound on the achievable rate. If the expressions for the bounds involve maximization with the respect to the joint distribution of random variables, it is assumed that the random variables are independent and uniformly distributed. For the fading relay channel, since the mutual information expressions which appear in the bounds are functions of the fading coefficients, expectation with respect to the fading coefficients is taken.

Results from \cite{HoZh}, where the capacity bounds and optimal power allocation strategies for the wireless FD and HD relay channel were studied, reveal that for the Gaussian input alphabet, the gain in rate offered by the relay channel over the direct transmission without the relay node, referred to as the relay gain, is significant and in particular at high SNR, this gain becomes a constant. 

The main contributions of the paper are as follows.
\begin{itemize}
\item We present lower and upper bounds on the achievable rate of the relay channel for both the  FD and HD scenarios.
\item It is shown that for the case of finite input constellations, the relay gain attains the  maximum at a particular SNR and at higher SNRs the relay gain tends to become zero. Since practical schemes always use finite input constellation, this  result means that the relay offers maximum advantage over the direct transmission when we operate at a particular SNR.  
\item The variation of the relay gain with the duty cycle of the HD relay channel is studied.
\end{itemize}

The organization of the paper is as follows. Section II describes the FD and the HD relay channel models. In Section III, we briefly discuss the already known capacity bounds for the relay channel and then we derive these bounds for input constellation constrained FD and HD  fading relay channels. In Section IV, the bounds and the relay gain with 4-QAM as the input constellation are computed. A comparison of  these results with already known results for the Gaussian input alphabet is made. Also, the variation of the relay gain with the duty cycle of the HD relay channel is studied.

\textit{\textbf{Notations:}} For a set $\cal S$,  $\mid$$\cal S$$\mid$ denotes  the cardinality of $\cal S.$ For a random variable $X_s$ which takes value from the set $\cal S$, we use $x_{s,i}$ to represent the $i$-th element of $\cal S$. $\textbf{E}_z[Y]$ denotes the expectation of $Y$ with respect to the random variable z. Throughout, $log$ refers to $log_2$ and $C(a)$ denotes $log(1+a)$.
 
\begin{figure}[htbp]
\centering
\includegraphics[totalheight=1in,width=3in]{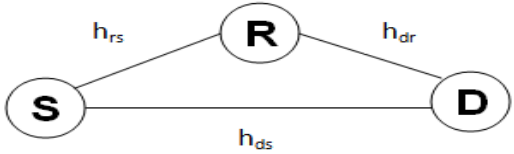}
\caption{The Relay Channel}	
\label{fig:Relay channel}	
\end{figure}
 
\section{CHANNEL MODEL}
\label{sec2}
We consider the relay channel shown in Fig. \ref{fig:Relay channel}, consisting of the source node $S$, the relay node $R$ and the destination node $D$. We focus on the Rayleigh fading scenario and let $h_{rs}$ = $c_{rs}$ $e^{j\phi_{rs}}$, $h_{ds}$ = $c_{ds}$ $e^{j\phi_{ds}}$ and $h_{dr}$ = $c_{dr}$ $e^{j\phi_{dr}}$ denote the zero mean complex Gaussian fading coefficients associated with the source-relay, source-destination and relay-destination links respectively, with the corresponding variances denoted by $\sigma_{rs}$,   $\sigma_{ds}$ and $\sigma_{dr}$. It is assumed that the power constraint at the source and the relay to be equal, denoted by $P$. Throughout, it is assumed that the global information about the channel state information (CSI) is available, i.e., all three nodes know the instantaneous values of $h_{rs}$, $h_{ds}$, $h_{dr}$. It is also assumed that all the nodes have perfect timing and carrier synchronization.

\begin{figure}[htbp]
\centering
\includegraphics[totalheight=1in,width=3in]{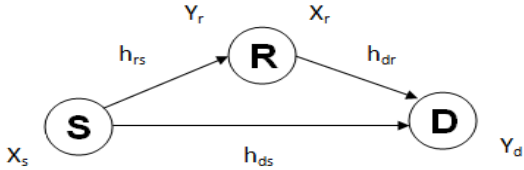}
\caption{Full Duplex Relay Channel}	
\label{fig:FD Relay channel}	
\end{figure}
\begin{figure}[htbp]
\centering
\includegraphics[totalheight=1in,width=3in]{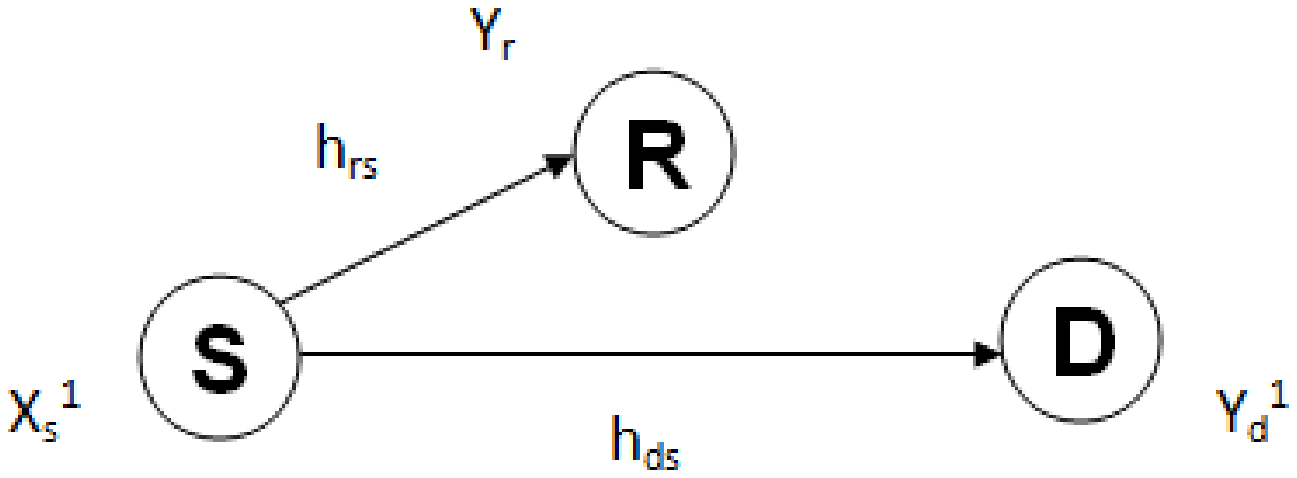}
\caption{HD Relay Channel - Phase 1}	
\label{fig:TD phase1}	
\end{figure}
\begin{figure}[htbp]
\centering
\includegraphics[totalheight=1in,width=3in]{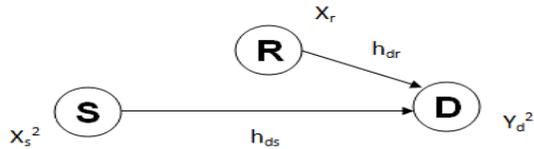}
\caption{HD Relay Channel - Phase 2}	
\label{fig:TD phase2}	
\end{figure}

The following two scenarios are considered: full-duplex (FD) relay channel (Fig. \ref{fig:FD Relay channel}), in which the relay node can receive and transmit simultaneously and the half-duplex (HD) relay channel, in which the relay receives during a fraction of the time $\alpha T$, and transmits for the duration $(1-\alpha) T$, where $T$ denotes the frame duration. In this paper we refer to $\alpha$ as the duty cycle of the relay. The receive and the transmit phases of the relay are referred to as Phase 1 (Fig. \ref{fig:TD phase1}) and Phase 2 (Fig. \ref{fig:TD phase2}) respectively. Since we have assumed that CSI is available at all the three nodes, the phases can be cancelled out at the appropriate places. The phase $\phi_{ds}$ due to the $S$-$D$ link can be cancelled out at S and the resulting phase $\phi_{rs}-\phi_{ds}$ at $R$ due to the $S$-$R$ link can be cancelled out at R. The phase $\phi_{dr}$ due to the $R$-$D$ link can be cancelled out at R. Hence we can take  $\phi_{rs}$=$\phi_{ds}$=$\phi_{dr}$=0.

For the FD relay the received signal at $R$ and $D$ are given by,
\begin {align}
\nonumber
Y_r &= c_{rs}X_s + z_r;\\
\nonumber
Y_d &= c_{ds}X_s + c_{dr}X_r + z_d.
\end {align}
\noindent \hspace{-.2 cm} where $X_s$ denotes the symbol transmitted by $S$, $z_r$ and $z_s$ are independent complex Gaussian random variables with mean 0 and variance 1/2 per dimension, denoted by $CG(0,1)$.

For the HD relay, the received signal at $R$ during Phase 1 is given by,
\begin {align}
\nonumber
\hspace{-1.5 cm} Y_r  = c_{rs}X_s^1 + z_r.
\end {align}
\noindent
The received signal at $D$ during Phase 1 and phase 2 are given by,
\begin {align}
\nonumber
&Y_d^1 = c_{ds}X_s^1 + z_d^1;\\
\nonumber
&Y_d^2 = c_{ds}X_s^2 + c_{dr}X_r + z_d^2.
\end {align}
\noindent 
where $X_s^1$ and $X_s^2$ denote the symbols transmitted by $S$ during Phase 1 and Phase 2 respectively, $X_r$ denotes the symbol transmitted by $R$ during Phase 2. $z_d^1$, $z_d^2$ and $z_r$ are independent and $CG(0,1)$. 

\section{BOUNDS ON THE ACHIEVABLE RATE}
\label{sec3}
The bounds on the achievable rate for fading relay channel with BPSK input were plotted in \cite{Turbo_full}\cite{Turbo_half}. In this section, the bounds for the fading relay channel with an arbitrary finite complex input constellation are derived. To derive the bounds, we follow a similar procedure as in \cite{HaR} \cite{AbR}.
\subsection{FD relay}

The lower and upper bounds on the capacity for the discrete memoryless FD relay channel \cite{CoEl} \cite{Cover} are given by,

{\footnotesize
\begin{align}
\label{LFDD}
C_{LFD}^{D}& =
max _{{P_{X_s,X_r}}}
min
\lbrace
I(X_s,X_r;Y_d),I(X_s;Y_r|X_r)
\rbrace;
\\
\label{UFDD}
C_{UFD}^{D}& = 
max _{{P_{X_s,X_r}}}
min
\lbrace
I(X_s,X_r;Y_d)],I(X_s;Y_r,Y_d|X_r)
\rbrace.
\end{align}
}
Throughout, it is assumed  that $X_s$ and $X_r$ are independent and hence the $max$ operation in \eqref{LFDD} and  \eqref{UFDD} is left out in the rest of the paper. For the FD relay with Gaussian alphabet, assuming that $X_s$ and $X_r$ are independent, the lower and upper bounds on the capacity are given by \cite{HoZh}

{\footnotesize
\begin{align}
\nonumber
&C_{LFD}^G = min
\left\lbrace
 \textbf{E}_{c_{ds},c_{dr}}[\textbf{$C$}\left((c_{ds}^2 + c_{dr}^2) P\right)],\textbf{E}_{c_{rs}}[\textbf{$C$}\left(c_{rs}^2 P\right)]
\right\rbrace;\\
\nonumber
&C_{UFD}^G = min
\lbrace
 \textbf{E}_{c_{ds},c_{dr}}[\textbf{$C$}\left((c_{ds}^2 + c_{dr}^2) P\right)],\\
 \nonumber
 & \hspace{4.7 cm} \textbf{E}_{c_{rs},c_{ds}}[\textbf{$C$}\left((c_{rs}^2+c_{ds}^2) P\right)]
\rbrace.
\end{align}
}

To compute the bounds on the achievable rate, it is assumed that $X_s$ and $X_ r$ are independent and uniformly distributed, with the same power $P$, $X_s$ $\in$ $S_s$, $X_r$ $\in$ $S_r$, with $\mid$$S_s$$\mid$ = $M_s$ and $\mid$$S_r$$\mid$ = $M_r.$

The achievable rate bounds of the FD relay channel with finite input constellation are given by,

{\small
\begin{align}
\label{ARFD}
C_{LFD}^{CC}& = min
\lbrace
\textbf{E}_{c_{ds},c_{dr}}[R_1],\textbf{E}_{c_{rs}}[R_2]
\rbrace;\\
C_{UFD}^{CC}& = min
\lbrace
\textbf{E}_{c_{ds},c_{dr}}[R_1],\textbf{E}_{c_{rs},c_{ds}}[R_3]
\rbrace.
\end{align}
}

Expressions for $R_1, R_2$ and $R_3$ are given in \eqref{r1}, \eqref{r2} and \eqref{r3} respectively along with their  explicit calculations in Appendix \ref{app-sec1}. The plot of the achievable rate bounds Vs P, for $\sigma_{ds}$ = -10 dB, $\sigma_{rs}$ = 2 dB and $\sigma_{dr}$ = 12 dB, for  the FD relay is shown in Fig. \ref{fig:FD_capacity} (in the next page). In Fig. \ref{fig:TD_capacity2} a similar plot is shown for $\sigma_{ds}$=-10 dB, $\sigma_{rs}$=12 dB and $\sigma_{dr}$=2 dB.

\subsection{HD relay}
For the discrete memoryless HD relay channel with duty cycle $\alpha$, the upper and lower bounds on the capacity are given by \cite{KhSaAa1},
{\small
\begin{align}
\nonumber
&C_{LHD}^D =
max _{{{P_{X_s^1,X_s^2,X_r}}}} 
 min
\lbrace
\alpha I(X_s^1;Y_r)\\
\nonumber
&\hspace{4.5 cm}+(1-\alpha)I(X_s^2;Y_d^2|X_r),\\
\label{LHDD}
&\hspace{2.8 cm} \alpha I(X_s^1;Y_d^1)+(1-\alpha)I(X_s^2,X_r;Y_d^2)
\rbrace;
\\
\nonumber
&C_{UHD}^D = 
max _{{P_{X_s^1,X_s^2,X_r}}}
min
\lbrace
\alpha I(X_s^1;Y_r,Y_d^1)\\
\nonumber
&\hspace{4.5 cm}+(1-\alpha)I(X_s^2;Y_d^2|X_r),\\
\label{UHDD}
&\hspace{2.8 cm} \alpha I(X_s^1;Y_d^1)]+(1-\alpha)I(X_s^2,X_r;Y_d^2)
\rbrace.
\end{align}
}
Similar to the FD relay, it is assumed  that $X_s^1$, $X_s^2$ and $X_r$ are independent and hence the $max$ operation in \eqref{LHDD} and \eqref{UHDD} are left out.

For the HD relay with Gaussian alphabet, assuming that $X_s^1$, $X_s^2$ and $X_r$ are independent, the lower and upper bounds on the capacity are given by \cite{KhSaAa2} \cite{HoZh},

{\small
\begin{align}
\nonumber
C_{LHD}^G& = min
\left\lbrace
\alpha \textbf{E}_{c_{rs}}\left[\textbf{$C$}\left(c_{rs}^2  P\right)\right]
+(1-\alpha) \textbf{E}_{c_{ds}}\left[\textbf{$C$}\left(c_{ds}^2  P\right)\right],\hspace{200 cm} \right\rbrace\\
\nonumber
&\hspace{-.2 cm} \hspace{-3 cm} \left\lbrace \hspace{2.2 cm}\alpha \textbf{E}_{c_{ds}}[\textbf{$C$}\left(c_{ds}^2 P\right)]+(1-\alpha)  \textbf{E}_{c_{ds},c_{dr}}\left[\textbf{$C$}\left(c_{ds}^2P+\dfrac{c_{dr}^2P}{1-\alpha}\right)\right]
\right\rbrace;\\
\nonumber
C_{UHD}^G& = min
\left\lbrace
\alpha \textbf{E}_{c_{ds},c_{rs}}[\textbf{$C$}\left((c_{ds}^2 + c_{rs}^2) P\right)]\hspace{200 cm} \right\rbrace\\
\nonumber
&\hspace{4 cm}+(1-\alpha)\textbf{E}_{c_{ds}}\left[\textbf{$C$}\left(c_{ds}^2 P\right)\right],\\
\nonumber
&\hspace{1.2 cm} (1-\alpha) \textbf{E}_{c_{ds},c_{dr}}\left[\textbf{$C$}\left(c_{ds}^2P + \dfrac{c_{dr}^2 P}{1-\alpha}\right)\right]\\
\nonumber
&\hspace{-.2 cm} \hspace{-3 cm} \left\lbrace \hspace{7 cm} +\alpha\textbf{E}_{c_{ds}}\left[\textbf{$C$}\left(c_{ds}^2 P\right)\right]
\right\rbrace.
\end{align}
}
Note that we have assumed the same power constraint P for both $S$ and $R$. $S$ uses the same power $P$ in both Phase 1 and Phase 2. Since the relay transmits only for a fraction (1-$\alpha$) of the total time, the transmit power $P$ of the relay has to be scaled by a factor 1/(1-$\alpha$).


To compute the achievable rate bounds with finite input constellation, assume $X_s^1$ $\in$ $S_s^1$, with $\mid$$S_s^1$$\mid$ = $M_{s_1}$, $X_s^2$ $\in$ $S_s^2$, with $\mid$$S_s^2$$\mid$ = $M_{s_2}$ and $X_r$ $\in$ $S_r$, with $\mid$$S_r$$\mid$ = $M_{r}$. 

Define $R_4$ = $I(X_s^1;Y_r)$, $R_5$ = $I(X_s^2;Y_d^2|X_r)$, $R_6$ = $I(X_s^1;Y_d^1)$, $R_7$ = $I(X_s^2,X_r;Y_d^2)$ and $R_8$ = $I(X_s^1;Y_r,Y_d^1)$.

 
The achievable rate bounds of the HD relay channel with finite input constellation are given by,

{\small
\begin{align}
\nonumber
&C_{LHD}^{CC} = min
\lbrace
\alpha\textbf{E}_{c_{rs}}[R_4]+(1-\alpha)\textbf{E}_{c_{ds}}[R_5],\\
\label{ARHD}
&\hspace{2.5 cm} \alpha\textbf{E}_{c_{ds}}[R_6]+(1-\alpha)\textbf{E}_{c_{ds},c_{dr}}[R_7]
\rbrace;\\
\nonumber
&C_{UHD}^{CC} = min
\lbrace
\alpha\textbf{E}_{c_{rs},c_{ds}}[R_8]+(1-\alpha)\textbf{E}_{c_{ds}}[R_5],\\
&\hspace{2.5 cm} \alpha\textbf{E}_{c_{ds}}[R_6]+(1-\alpha)\textbf{E}_{c_{ds},c_{dr}}[R_7]
\rbrace.
\end{align}
}

Expressions for $R_4, R_5, R_6, R_7$ and $R_8$ are given in \eqref{r4}-\eqref{r8} along with their explicit calculations in Appendix \ref{app-sec2}. The plot of the achievable rate bounds Vs P for the HD relay, for $\sigma_{ds}$=-10 dB, $\sigma_{rs}$=2 dB, $\sigma_{dr}$=12 dB and $\alpha$=0.5 is shown in Fig. \ref{fig:TD_capacity} (in the next page). In Fig. \ref{fig:TD_capacity2} a similar plot is shown for $\sigma_{ds}$=-10 dB, $\sigma_{rs}$=12 dB, $\sigma_{dr}$=2 dB and $\alpha$=0.5.
\subsection{Direct transmission}
Our goal is to compare the bounds for FD and HD relay with the capacity of direct transmission without the relay and in order to make the comparison fair, we assume that the power constraint at $S$ is $2P$.
For direct transmission with Gaussian input alphabet, the capacity is given by,
\begin{align}
\nonumber
C_{DIR}^G = \textbf{E}_{c_{ds}}[C(2 c_{ds}^2 P)]
\end{align}
\begin{figure*} 
\scriptsize
\begin{align}
\label{r9}
&R_9~\triangleq~I(X_s;Y_d)]=log(M_s)-\frac{1}{M_s}\sum_{i_1=0}^{M_s-1}\textbf{E}_{z_d}\left[log\left(\dfrac{\sum_{i = 0}^{M_s-1}exp\left(-\vert z_d-c_{ds}x_{s,i}+c_{ds}x_{s,i_1}\vert ^2\right)}{exp(-\vert z_d \vert^2)}\right)\right]
\end{align}
\end{figure*}
and to compute the CCC with finite input constellation, assume $X_s$ $\in$ $S_s$, $\mid$$S_s$$\mid$ = $M_s$ and define $R_9$~=~$I(X_s;Y_d)$. $R_9$ can be computed and is given by \eqref{r9} (in the next page). The CCC for the direct transmission without the relay, with finite input constellation is given by, 
\begin{align}
\label{ARD}
C_{DIR}^{CC} = \textbf{E}[R_9]. 
\end{align}
Fig. \ref{fig:FD_capacity} and Fig. \ref{fig:TD_capacity} show a comparison of $C_{DIR}^{CC}$ with the achievable rate bounds of the FD and the HD relay channels.

\section{RELAY GAIN COMPUTATION}
\begin{definition}
Relay gain is defined as the difference between the lower bound on the achievable rate in the presence of the relay and the capacity of the direct transmission without the relay.
\end{definition}

In this section we compute the achievable rate bounds and the relay gain for the FD and HD relay channels. The capacity bounds and relay gain for the fading FD and HD relay channels with Gaussian input alphabet were studied in \cite{HoZh}. It is assumed that the finite input constellation used at S and R are the same, i.e., $M_s$ = $M_r$ = $M$ for the FD relay and $M_{s_1}$ = $M_{s_2}$ = $M_r$ = $M$ for the HD relay.

\begin{figure}[htbp]
\centering
\includegraphics[totalheight=2.5in,width=3.5in]{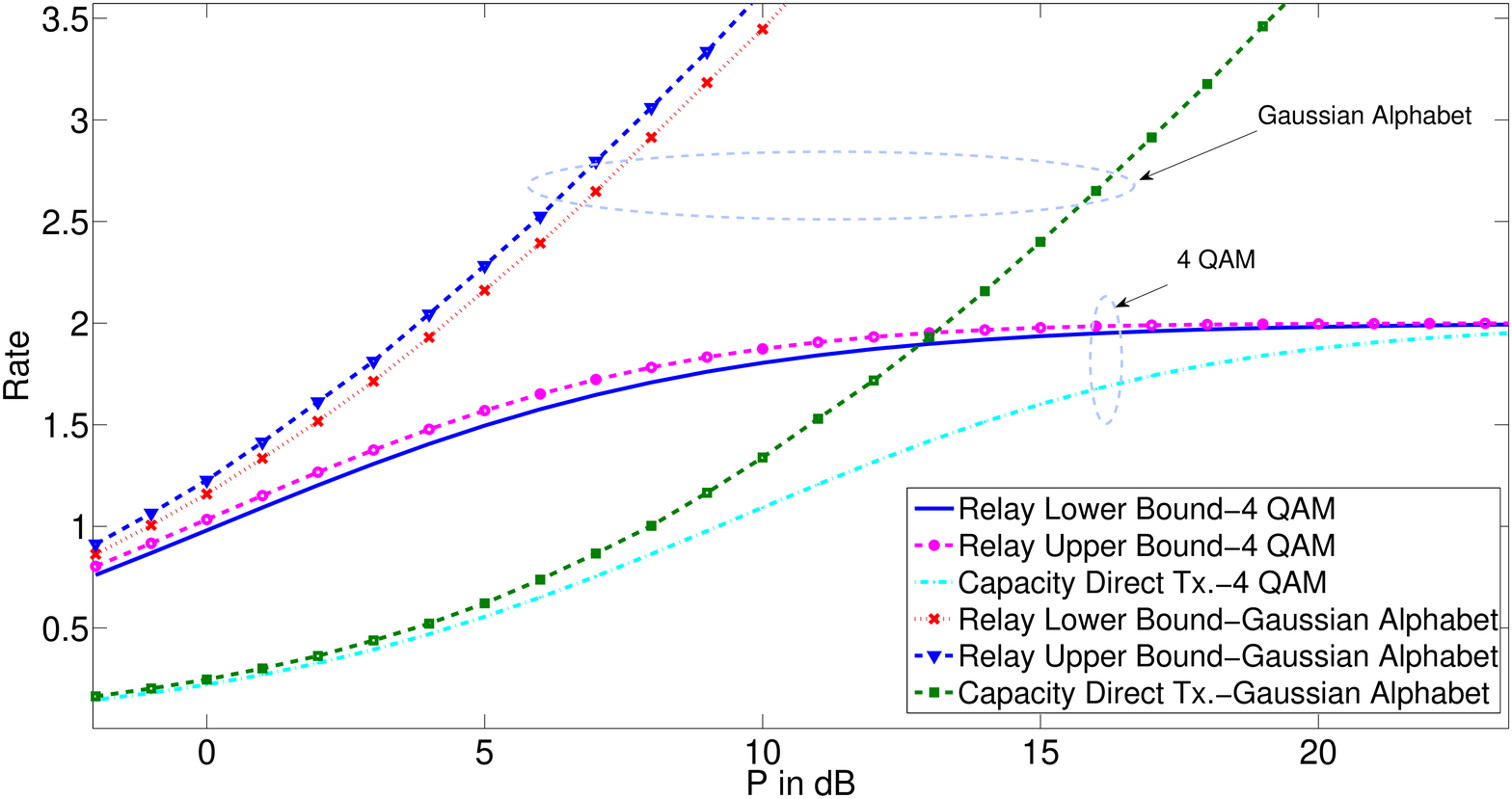}
\caption{Rate Vs P for FD relay with $\sigma_{ds}$=-10 dB, $\sigma_{rs}$=2 dB, $\sigma_{dr}$=12 dB \hspace{200 cm}}	
\label{fig:FD_capacity}	
\end{figure}

\begin{figure}[htbp]
\centering
\includegraphics[totalheight=2.5in,width=3.5in]{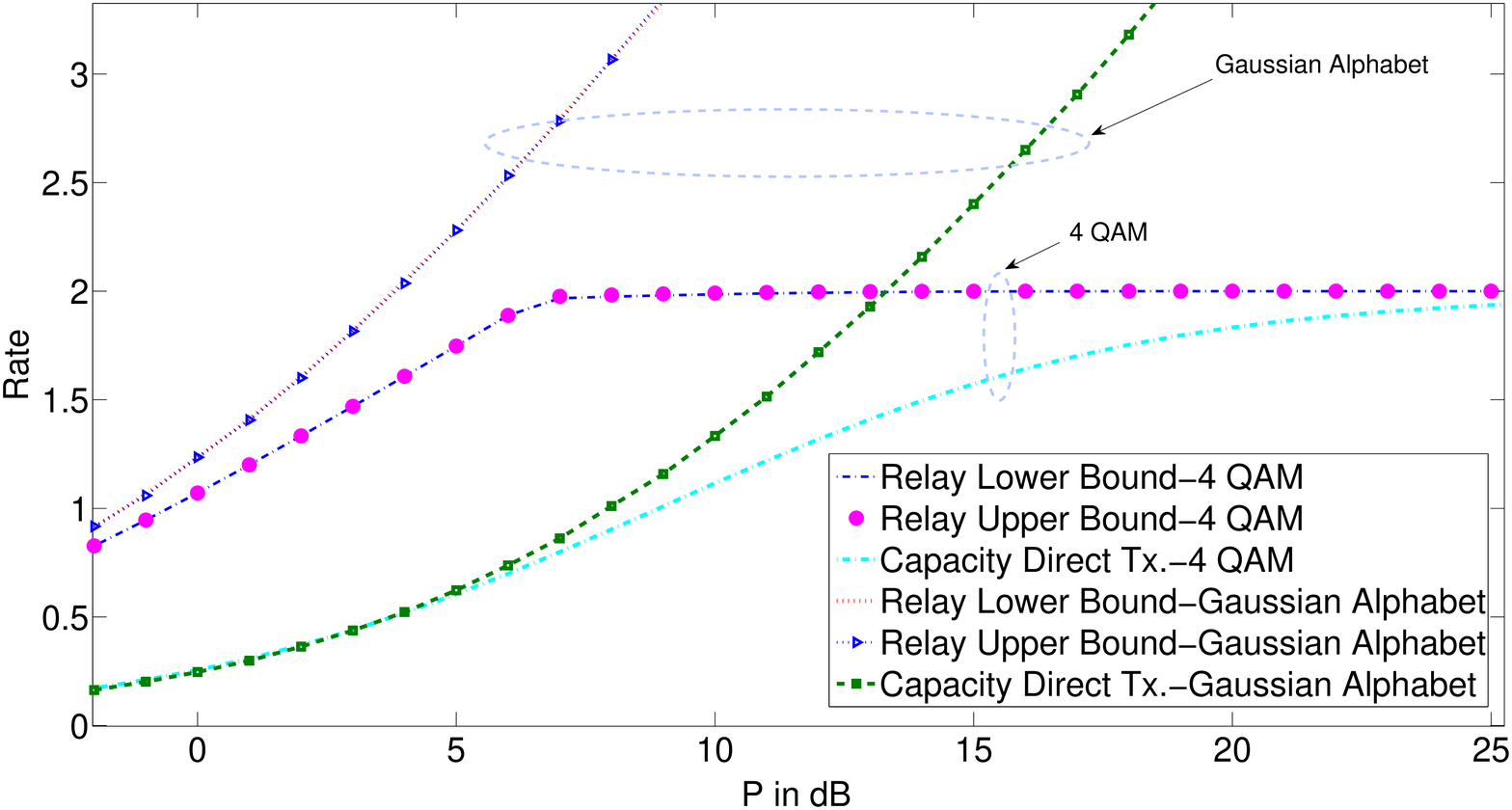}
\caption{Rate Vs P for FD relay with $\sigma_{ds}$=-10 dB, $\sigma_{rs}$=12 dB, $\sigma_{dr}$=2 dB \hspace{200 cm}}	
\label{fig:FD_capacity2}	
\end{figure}

\begin{figure}[htbp]
\centering
\includegraphics[totalheight=2.5in,width=3.5in]{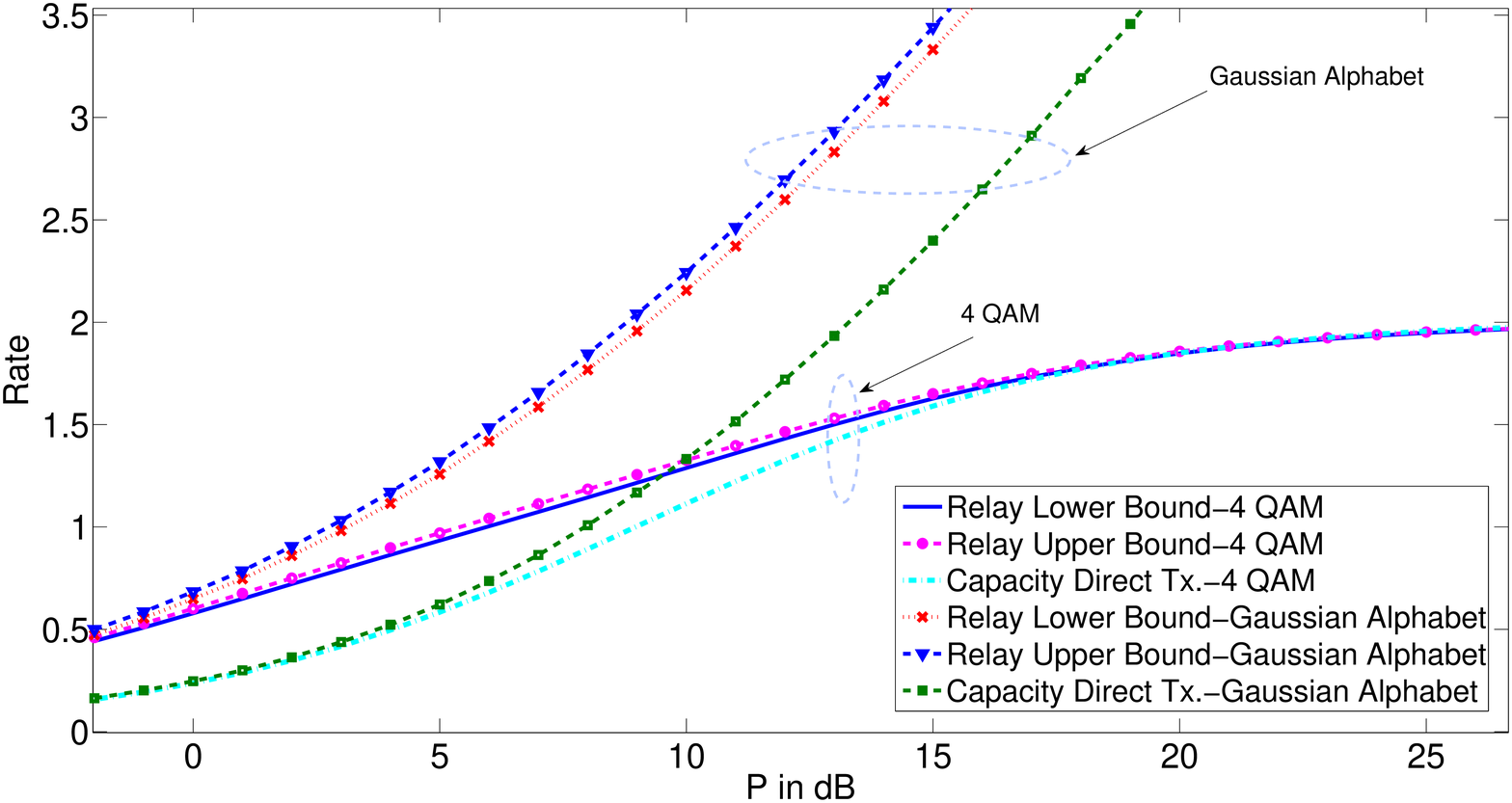}
\caption{Rate Vs P for HD relay with $\sigma_{ds}$=-10 dB, $\sigma_{rs}$=2 dB, $\sigma_{dr}$=12 dB, $\alpha$=0.5}	
\label{fig:TD_capacity}	
\end{figure}

\begin{figure}[htbp]
\centering
\includegraphics[totalheight=2.5in,width=3.5in]{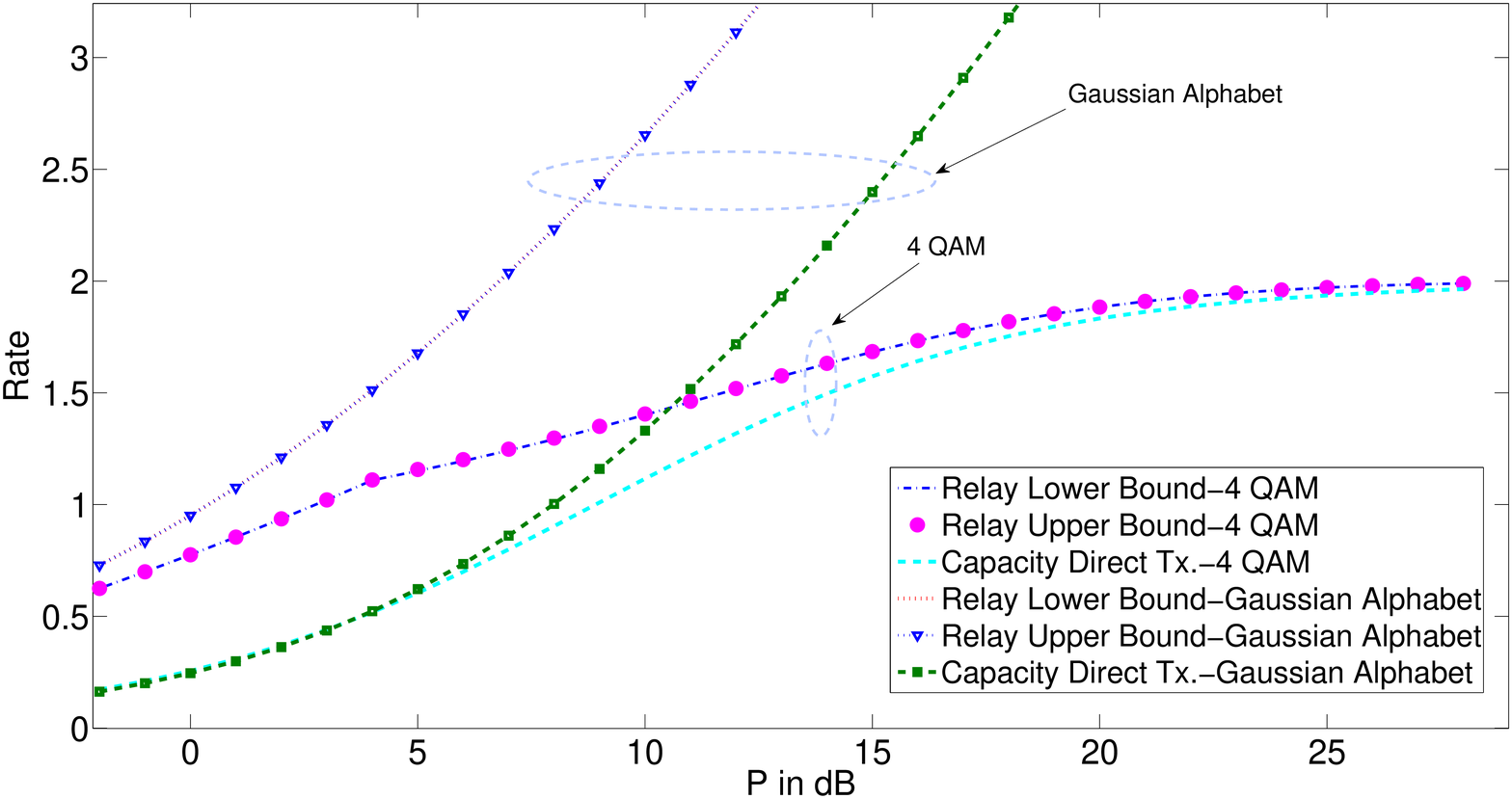}
\caption{Rate Vs P for HD relay with $\sigma_{ds}$=-10 dB, $\sigma_{rs}$=12 dB, $\sigma_{dr}$=2 dB, $\alpha$=0.5}	
\label{fig:TD_capacity2}	
\end{figure}

The expressions for the capacity bounds, given in Appendix \ref{app-sec1}  and Appendix \ref{app-sec2}, can be evaluated using Monte-Carlo simulations. Fig. \ref{fig:FD_capacity} and Fig. \ref{fig:FD_capacity2} show the Rate Vs P plot for the FD relay channel, showing the lower and upper bounds on the capacity with Gaussian and 4-QAM input alphabets, as well as the capacity of the direct transmission with Gaussian and 4-QAM input alphabets, for  two different sets of $\sigma_{ds}$, $\sigma_{rs}$ and $\sigma_{dr}$. In  Fig. \ref{fig:TD_capacity} and \ref{fig:TD_capacity2}, similar plots are shown for the HD relay with $\alpha$=0.5.

From Fig. \ref{fig:FD_capacity}, Fig. \ref{fig:FD_capacity2}, Fig. \ref{fig:TD_capacity} and Fig. \ref{fig:TD_capacity2}, under the given assumptions, we observe that the upper and lower bounds on the capacity are very close, not only for  the Gaussian input alphabet, consistent with the results in \cite{HoZh}, but also for the finite input constellation (4-QAM). At low values of P, the rate achievable with the use of finite input constellation is nearly the same as the rate achievable using Gaussian input alphabet.  

\subsection{Variation of relay gain with P}
\begin{figure}[htbp]
\centering
\includegraphics[totalheight=2in,width=3.5in]{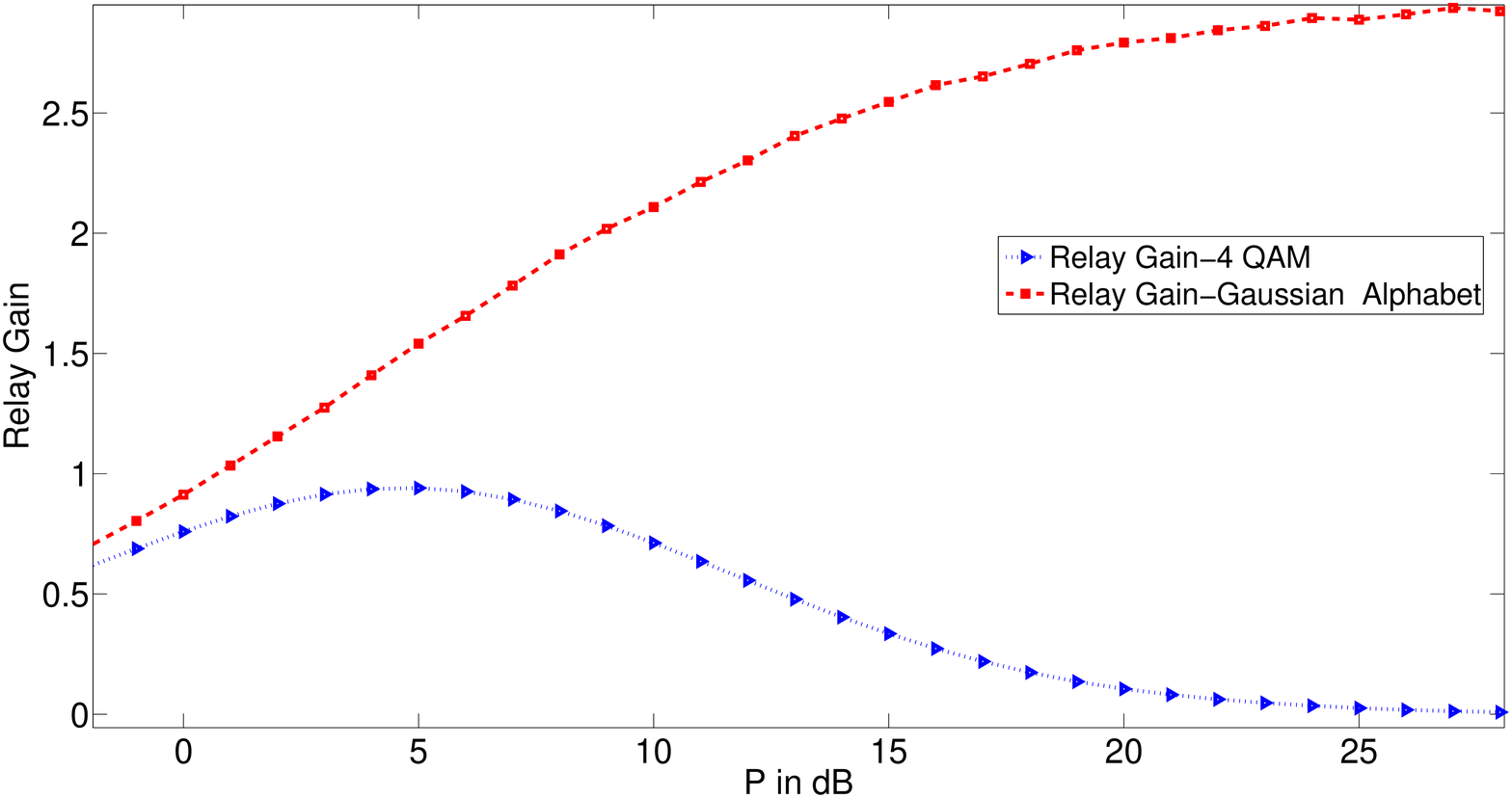}
\caption{Relay Gain Vs P  for FD relay with $\sigma_{ds}$=-10 dB, $\sigma_{rs}$=2 dB, $\sigma_{dr}$=12 dB}	
\label{fig:FD_gain}	
\end{figure}

\begin{figure}[htbp]
\centering
\includegraphics[totalheight=2in,width=3.5in]{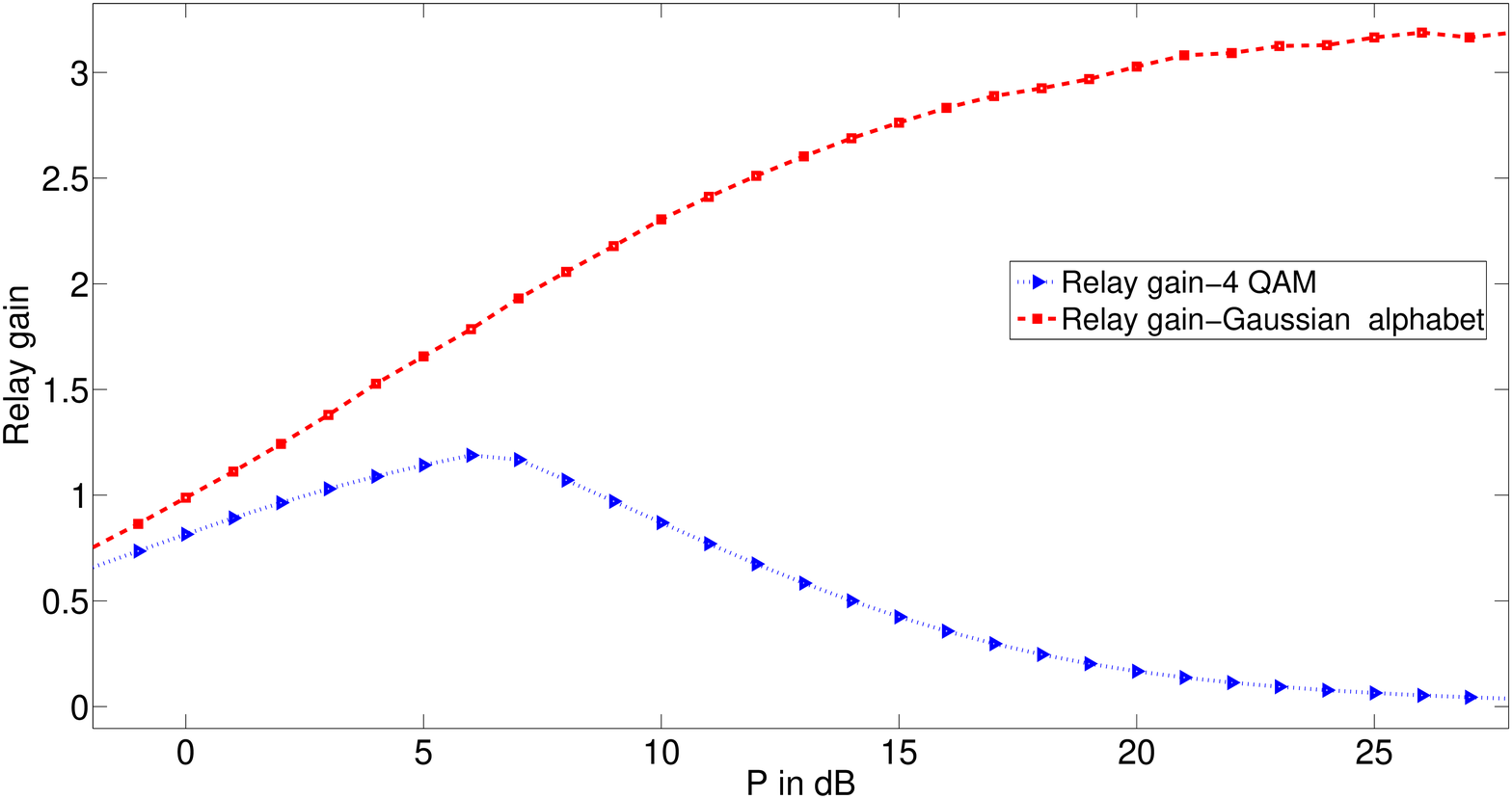}
\caption{Relay Gain Vs P  for FD relay with $\sigma_{ds}$=-10 dB, $\sigma_{rs}$=12 dB, $\sigma_{dr}$=2 dB}	
\label{fig:FD_gain2}	
\end{figure}

\begin{figure}[htbp]
\centering
\includegraphics[totalheight=2in,width=3.5in]{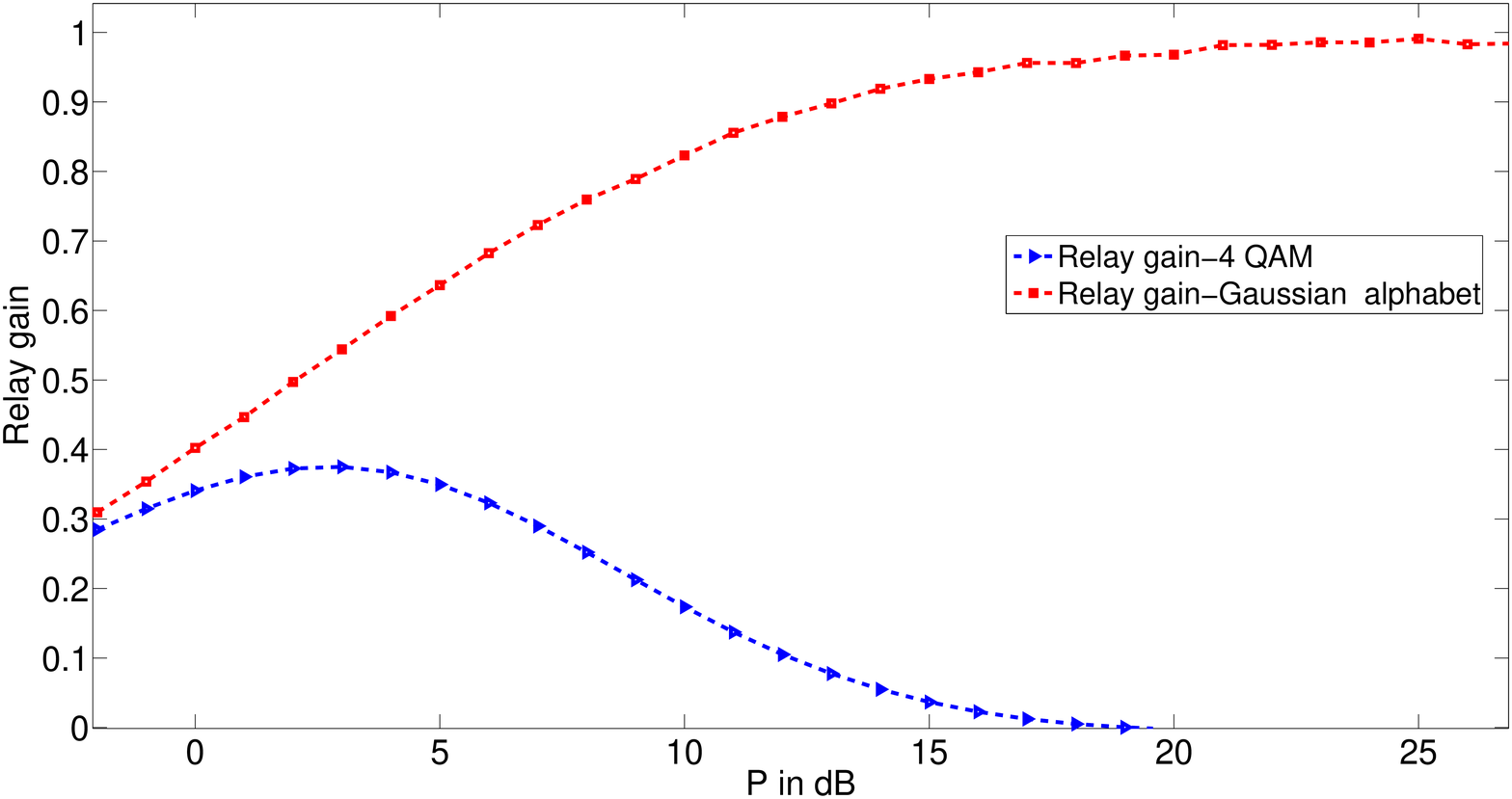}
\caption{Relay Gain Vs P for HD relay with $\sigma_{ds}$=-10 dB, $\sigma_{rs}$=2 dB, $\sigma_{dr}$=12 dB, $\alpha$=0.5}	
\label{fig:TD_gain}	
\end{figure}

\begin{figure}[htbp]
\centering
\includegraphics[totalheight=2in,width=3.5in]{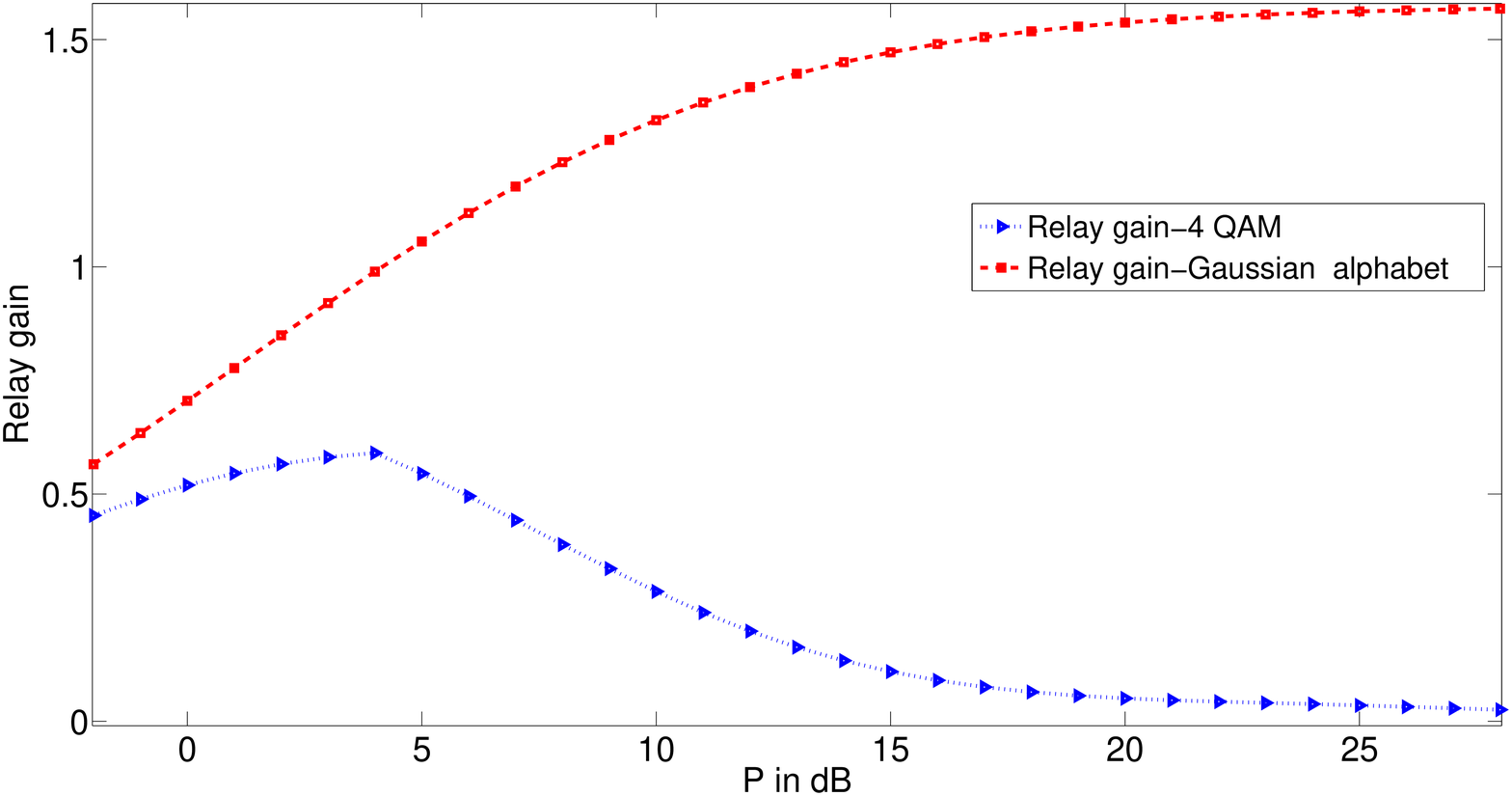}
\caption{Relay Gain Vs P for HD relay with $\sigma_{ds}$=-10 dB, $\sigma_{rs}$=12 dB, $\sigma_{dr}$=2 dB, $\alpha$=0.5}	
\label{fig:TD_gain2}	
\end{figure}

The relay gain, which is the difference between the curves corresponding to the achievable rate in presence of relay and the capacity of the direct transmission, in Fig. \ref{fig:FD_capacity}, Fig. \ref{fig:FD_capacity2}, Fig. \ref{fig:TD_capacity} and Fig. \ref{fig:TD_capacity2}, is plotted in Fig. \ref{fig:FD_gain}, Fig. \ref{fig:FD_gain2}, Fig. \ref{fig:TD_gain} and Fig. \ref{fig:TD_gain2}. As observed in \cite{HoZh}, with Gaussian input alphabet, the relay gain increases with P initially and at high values of P, becomes a constant. Expressions for the asymptotic relay gain (relay gain as P $\rightarrow$ $\infty$) for the FD and HD relay with Gaussian input alphabet were derived in \cite{HoZh}.

Contrarily, with finite input constellation the asymptotic relay gain is always zero. Assuming that the finite constellation used at S and R are of the same cardinality $M$, it can be easily seen from the equation sets $\lbrace$\eqref{ARFD},\eqref{r1},\eqref{r2}$\rbrace$, $\lbrace$\eqref{r4}-\eqref{r7},\eqref{ARHD}$\rbrace$ and $\lbrace$\eqref{r9},\eqref{ARD}$\rbrace$ that as P $\rightarrow$ $\infty$ , the achievable rate with the FD and HD relays as well as the capacity of the direct transmission become equal to $log(M)$ and hence the asymptotic relay gain is zero. As seen in Fig. \ref{fig:FD_gain}, Fig. \ref{fig:FD_gain2}, Fig. \ref{fig:TD_gain} and Fig. \ref{fig:TD_gain2}, the relay gain with finite input constellation, attains a peak value at a particular value of P and becomes zero at high values of P. The same argument holds good for arbitrary values of $\sigma_{ds}$, $\sigma_{rs}$ and $\sigma_{dr}$. This peak value of the relay gain is referred to as the \textit{maximum relay gain}.

It is interesting to note that similar conclusions were observed for the secrecy capacity of the wire-tap channel with Gaussian and M-PAM input alphabets in \cite{MiRoAn} and with QAM, PSK input alphabets in \cite{RaR}.

\begin{figure}[htbp]
\centering
\includegraphics[totalheight=2in,width=3.5in]{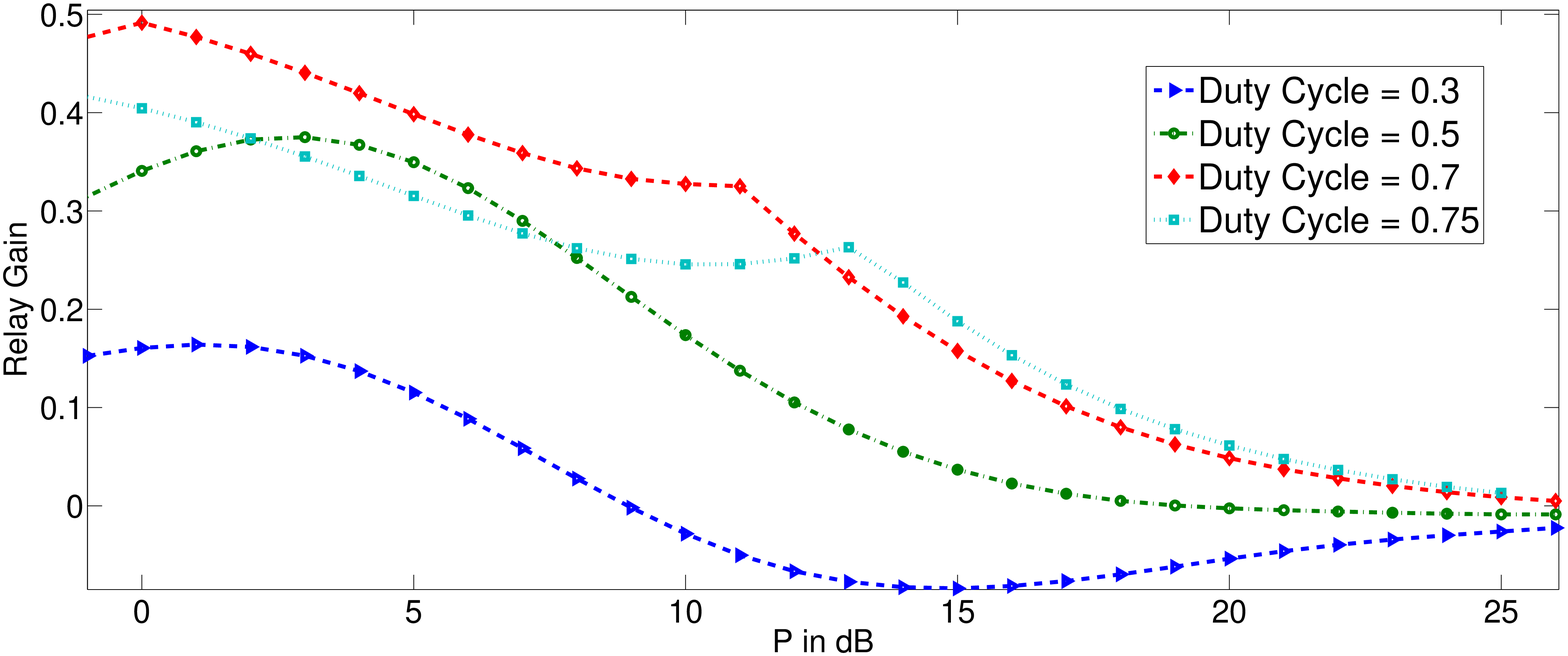}
\caption{Relay Gain for HD relay for 4-QAM with $\sigma_{ds}$=-10 dB, $\sigma_{rs}$=2 dB, $\sigma_{dr}$=12 dB for four different values of $\alpha$}	
\label{fig:TD_duty_cycle}	
\end{figure}
\begin{figure}[htbp]
\centering
\includegraphics[totalheight=2in,width=3.5in]{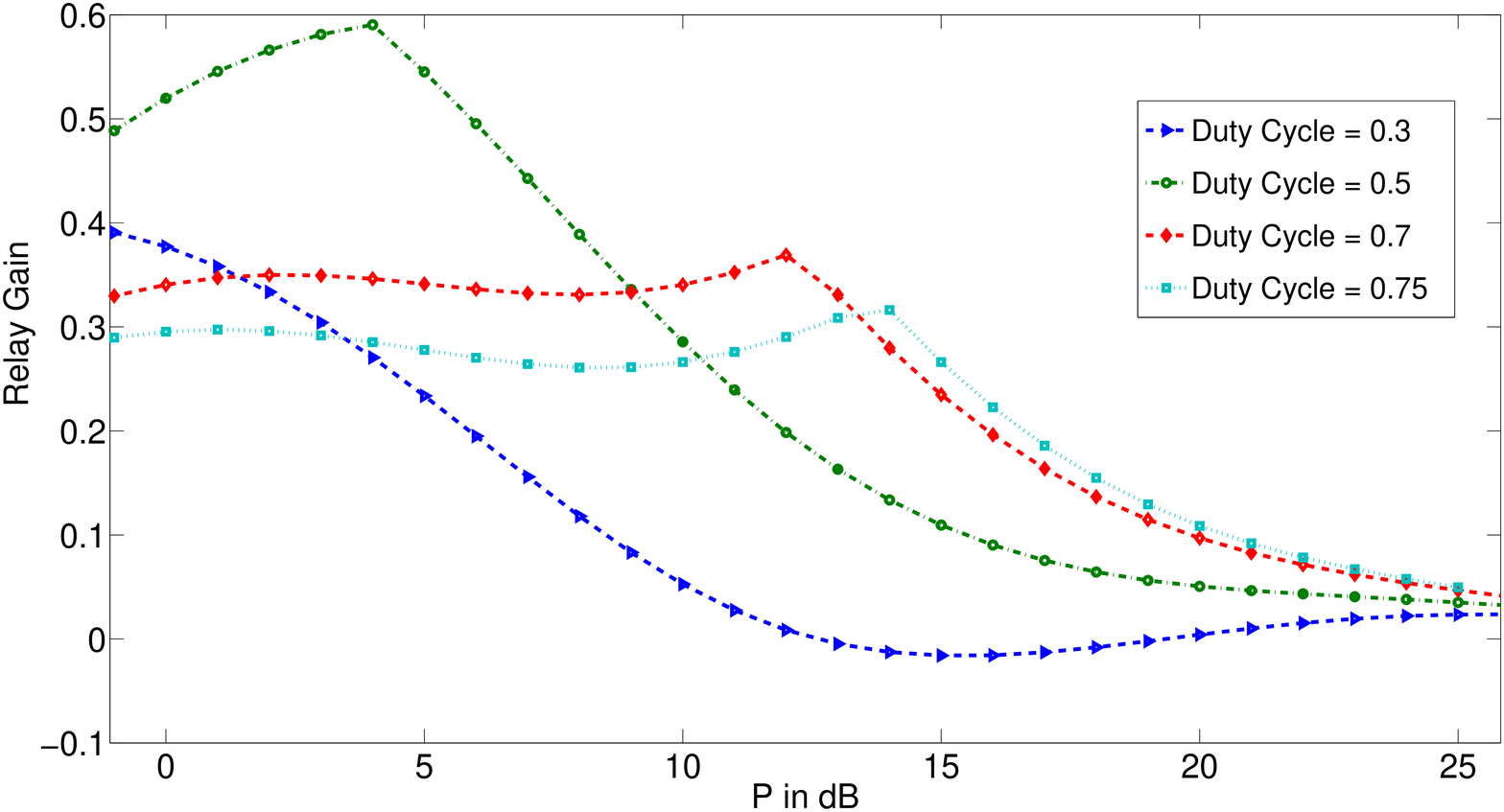}
\caption{Relay Gain for HD relay for 4-QAM with $\sigma_{ds}$=-10 dB, $\sigma_{rs}$=12 dB, $\sigma_{dr}$=2 dB for four different values of $\alpha$}	
\label{fig:TD_duty_cycle2}	
\end{figure}
\begin{figure}[htbp]
\centering
\includegraphics[totalheight=2in,width=3.5in]{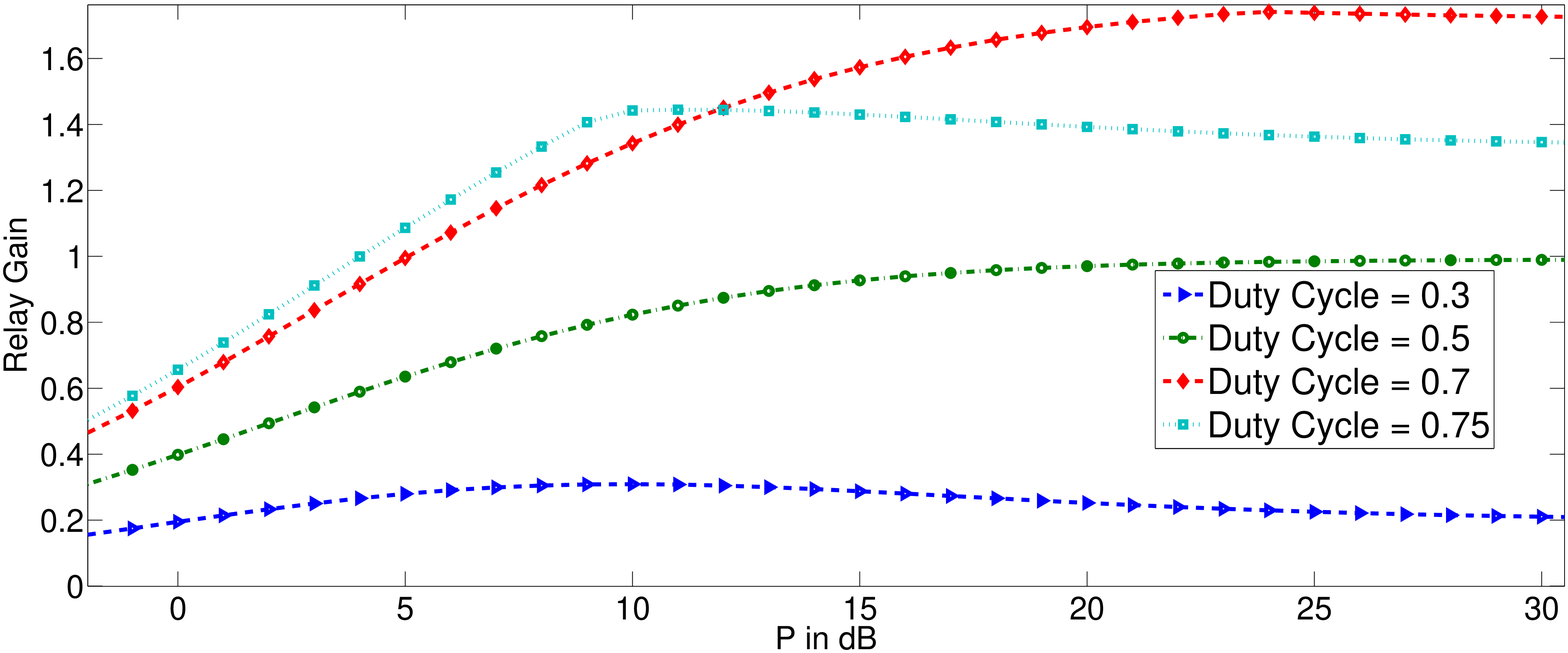}
\caption{Relay Gain for HD relay for Gaussian alphabet with $\sigma_{ds}$=-10 dB, $\sigma_{rs}$=2 dB, $\sigma_{dr}$=12 dB for four different values of $\alpha$}	
\label{fig:gaussian_duty_cycle}	
\end{figure}
\begin{figure}[htbp]
\centering
\includegraphics[totalheight=2in,width=3.5in]{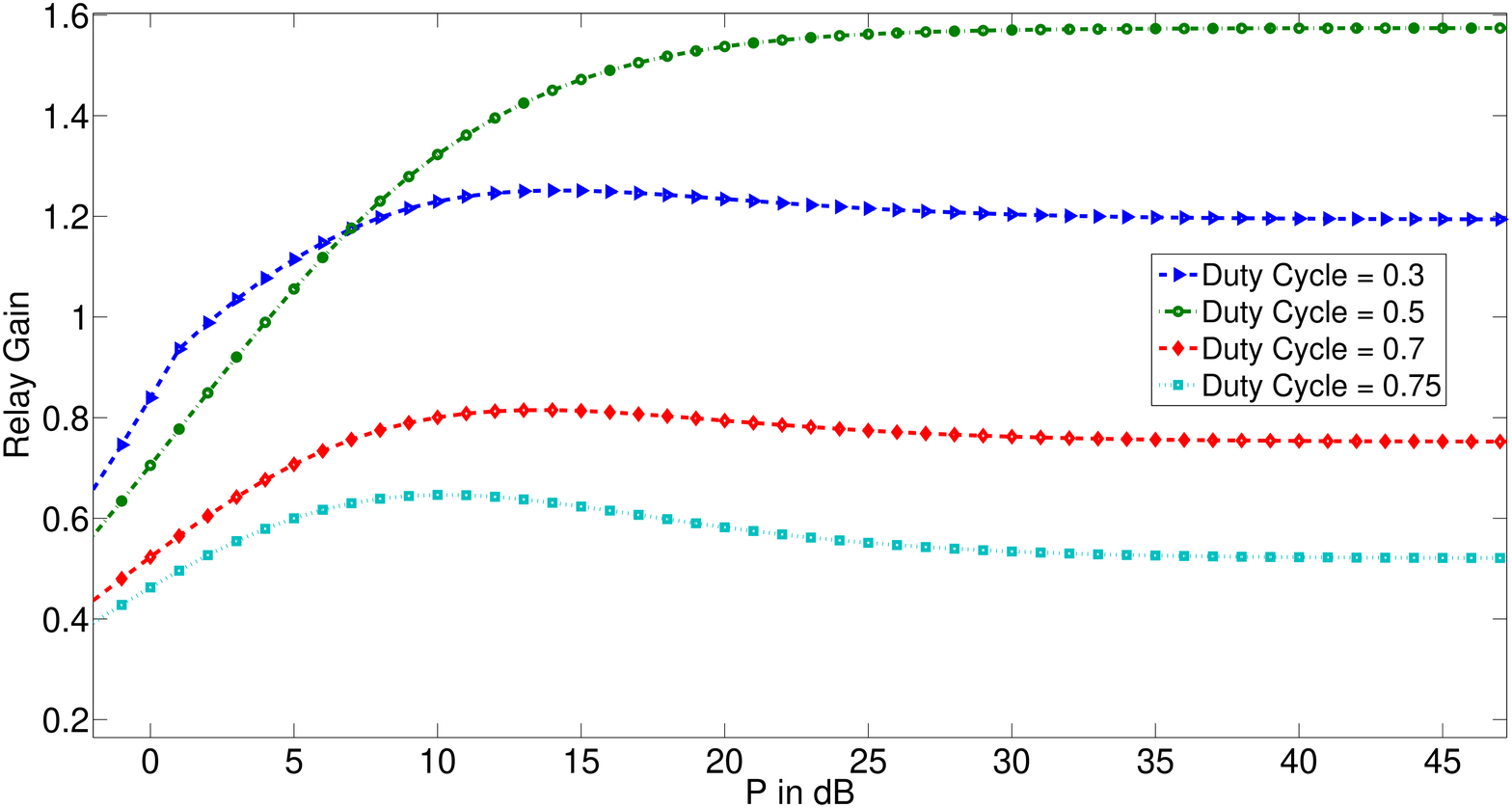}
\caption{Relay Gain for HD relay for Gaussian alphabet with $\sigma_{ds}$=-10 dB, $\sigma_{rs}$=12 dB, $\sigma_{dr}$=2 dB for four different values of $\alpha$}	
\label{fig:gaussian_duty_cycle2}	
\end{figure}

\subsection{Variation of Relay gain for HD relay with $\alpha$}
Fig. \ref{fig:TD_duty_cycle} shows the Relay gain Vs P plots for four different values of $\alpha$, for 4-QAM input alphabet, with $\sigma_{ds}$ = -10 dB, $\sigma_{rs}$ = 2 dB and $\sigma_{dr}$ = 12 dB. As seen from Fig. \ref{fig:TD_duty_cycle}, the maximum relay gain increases when $\alpha$ increases from 0.3 to 0.7. When $\alpha$ increases from 0.7 to 0.75, the maximum relay gain decreases. This means that the maximum relay gain as a function of $\alpha$, attains a peak for some $\alpha$ between 0.7 and 0.75. Similarly, from Fig. \ref{fig:TD_duty_cycle2} we can see that for $\sigma_{ds}$ = -10 dB, $\sigma_{rs}$ = 12 dB and $\sigma_{dr}$ = 2 dB, the maximum relay gain as a function of $\alpha$, attains a peak for some $\alpha$ between 0.3 and 0.5.

Fig. \ref{fig:gaussian_duty_cycle} shows the Relay Gain Vs P plots for four different values of $\alpha$, for Gaussian input alphabet, with $\sigma_{ds}$ = -10 dB, $\sigma_{rs}$ = 2 dB and $\sigma_{dr}$ = 12 dB. As seen from Fig. \ref{fig:gaussian_duty_cycle}, the asymptotic relay gain increases when $\alpha$ increases from 0.3 to 0.7. When $\alpha$ increases from 0.7 to 0.75, the asymptotic relay gain decreases. The asymptotic relay gain as a function of $\alpha$, attains a peak for some $\alpha$ between 0.7 and 0.75. Similarly, from Fig. \ref{fig:gaussian_duty_cycle2} we can see that for $\sigma_{ds}$ = -10 dB, $\sigma_{rs}$ = 12 dB and $\sigma_{dr}$ = 2 dB, the asymptotic relay gain as a function of $\alpha$, attains a peak for some $\alpha$ between 0.3 and 0.5.

From Fig. \ref{fig:TD_duty_cycle}, Fig. \ref{fig:TD_duty_cycle2}, Fig. \ref{fig:gaussian_duty_cycle} and Fig. \ref{fig:gaussian_duty_cycle2}, it is very clear that the variation of the relay gain with $P$ and $\alpha$ is not the same with Gaussian alphabet and 4-QAM. The fact that with 4-QAM, the maximum advantage over the direct transmission is obtained at a particular value of $P$ and $\alpha$ cannot be seen from the relay gain plots for Gaussian input alphabet (Fig. \ref{fig:gaussian_duty_cycle} and Fig. \ref{fig:gaussian_duty_cycle2}).  
\section{DISCUSSION}
It is shown that the variation of relay gain as a function of SNR with finite input constellation is different from the variation with continuous Gaussian input assumption. It will be  interesting to study how the Relay gain Vs SNR curves vary with increasing size of the finite input constellation. Further it will be interesting to study the channel conditions under which the relay gain with finite input constellation is negative, i.e., the use of relay offers no advantage over the direct transmission without the relay. 

\section*{Acknowledgement}
We thank G. Abhinav for the useful discussions. This work was supported  partly by the DRDO-IISc program on Advanced Research in Mathematical Engineering through a research grant as well as the INAE Chair Professorship grant to B.~S.~Rajan.

\newpage

\appendices

\section{Calculation of bounds for the FD relay channel with finite input constellations}
\label{app-sec1}

\begin{figure*}
\scriptsize
\begin{align}
\label{density1}
&P(Y_d=y_d|X_s=x_{s,i},X_r=x_{r,j}) = \frac{1}{\pi} \exp\left(-\vert y_d-c_{ds}x_{s,i}-c_{dr}x_{r,j}\vert ^2\right)\\
\label{density2}
&P(Y_d=y_d) = \dfrac{1}{M_sM_r}\sum_{i = 0}^{M_s-1}\sum_{j = 0}^{M_r-1}P(Y_d =y_d|X_s=x_{s,i},X_r=x_{r,j})\\
\label{density3}
&P(Y_r=y_r|X_s=x_{s,i},X_r=x_{r,j}) = \frac{1}{\pi} \exp\left(-\vert y_r-c_{rs}x_{s,i}\vert ^2\right)\\
\label{density4}
&P(Y_r=y_r|X_r=x_{r,j})=\frac{1}{M_s}\sum_{i = 0}^{M_s-1}P(Y_r=y_r|X_s=x_{s,i},X_r=x_{r,j})\\
\label{density5}
&P(Y_r=y_r,Y_d=y_d|X_s=x_{s,i},X_r=x_{r,j}) = \frac{1}{\pi^2} \exp\left(-\vert y_r-c_{rs}x_{s,i}\vert ^2-\vert y_d-c_{ds}x_{s,i}-c_{dr}x_{r,j}\vert ^2\right)\\
\label{density6}
&P(Y_r=y_r,Y_d=y_d|X_r=x_{r,j})=\frac{1}{M_s}\sum_{i = 0}^{M_s-1}P(Y_r=y_r,Y_d=y_d|X_s=x_{s,i},X_r=x_{r,j})\\
\hline
\label{h1}
&h(Y_d)=log(\pi M_sM_r)-\dfrac{1}{M_sM_r}\sum_{i_1 =0}^{M_s-1}\sum_{j_1 = 0}^{M_r-1}\textbf{E}_{z_d}\left[log\left({\sum_{i = 0}^{M_s-1}\sum_{j =0}^{M_r-1}exp\left(-\vert z_d-c_{ds}x_{s,i}-c_{dr}x_{r,j}+c_{ds}x_{s,i_1}+c_{dr}x_{r,j_1}\vert ^2\right)}\right)\right]\\
\label{h2}
&h(Y_d|X_s,X_r)=log(\pi)-\dfrac{1}{M_sM_r}\sum_{i_1 =0}^{M_s-1}\sum_{j_1 = 0}^{M_r-1}\textbf{E}_{z_d}\left[log\left(exp(-\vert z_d \vert^2)\right)\right]\\
\nonumber
&R_1~\triangleq~I(X_s,X_r;Y_d)=log(M_sM_r)\\  
\label{r1}
& \hspace{3.5 cm} -\dfrac{1}{M_sM_r}\sum_{i_1 =0}^{M_s-1}\sum_{j_1 = 0}^{M_r-1}\textbf{E}_{z_d}\left[log\left(\dfrac{\sum_{i = 0}^{M_s-1}\sum_{j =0}^{M_r-1}exp\left(-\vert z_d-c_{ds}x_{s,i}-c_{dr}x_{r,j}+c_{ds}x_{s,i_1}+c_{dr}x_{r,j_1}\vert ^2\right)}{exp(-\vert z_d \vert^2)}\right)\right]\\
\hline
\label{h3}
&h(Y_r|X_r,X_s)=log(\pi M_s)
-\frac{1}{M_s}\sum_{i_1=0}^{M_s-1}\textbf{E}_{z_r}\left[log\left({exp(-\vert z_r \vert^2)}\right)\right]\\
\label{h4}
&h(Y_r|X_r)]=log(\pi M_s)
-\frac{1}{M_s}\sum_{i_1=0}^{M_s-1}\textbf{E}_{z_r}\left[log\left({\sum_{i = 0}^{M_s-1}exp\left(-\vert z_r-c_{rs}x_{s,i}+c_{rs}x_{s,i_1}\vert ^2\right)}\right)\right]\\
\label{r2}
&R_2~\triangleq~I(X_s;Y_r|X_r)]=log(M_s)
-\frac{1}{M_s}\sum_{i_1=0}^{M_s-1}\textbf{E}_{z_r}\left[log\left(\dfrac{\sum_{i = 0}^{M_s-1}exp\left(-\vert z_r-c_{rs}x_{s,i}+c_{rs}x_{s,i_1}\vert ^2\right)}{exp(-\vert z_r \vert^2)}\right)\right]\\
\hline
\label{h5}
&h(Y_r,Y_d|X_r)=log(\pi^2 M_s)-\frac{1}{M_s}\sum_{i_1=0}^{M_s-1}\textbf{E}_{z_r ,z_d}\left[log\left({\sum_{i = 0}^{M_s-1}exp\left(-\vert z_r-c_{rs}x_{s,i}+c_{rs}x_{s,i_1}\vert ^2-\vert z_d-c_{ds}x_{s,i}+c_{ds}x_{s,i_1}\vert ^2\right)}\right)\right]\\
\label{h6}
&h(Y_r,Y_d|X_r,X_s)=log(\pi^2 M_s)-\frac{1}{M_s}\sum_{i_1=0}^{M_s-1}\textbf{E}_{z_r ,z_d}\left[log\left({exp(-\vert z_r \vert^2-\vert z_d \vert^2)}\right)\right]\\
\label{r3}
&R_3~\triangleq~I(X_s;Y_r,Y_d|X_r)]=log(M_s)-\frac{1}{M_s}\sum_{i_1=0}^{M_s-1}\textbf{E}_{z_r ,z_d}\left[log\left(\dfrac{\sum_{i = 0}^{M_s-1}exp\left(-\vert z_r-c_{rs}x_{s,i}+c_{rs}x_{s,i_1}\vert ^2-\vert z_d-c_{ds}x_{s,i}+c_{ds}x_{s,i_1}\vert ^2\right)}{exp(-\vert z_r \vert^2-\vert z_d \vert^2)}\right)\right]\\
\hline
\nonumber
\end{align}
\end{figure*}

Define $R_1$~=~$I(X_s,X_r;Y_d)$, $R_2$~=~$I(X_s;Y_r|X_r)$, and $R_3$~=~$I(X_s;Y_r,Y_d|X_r)$.
\subsubsection{Calculation of $R_1$}
For the FD relay,
$I(X_s,X_r;Y_d)$ is given by,
{\small
\begin {align}
\nonumber
I(X_s,X_r;Y_d) = h(Y_d)-h(Y_d|X_s,X_r), 
\end {align}
}
\hspace{-0.25 cm} where $h(Y_d)$ can be calculated using the density $P(Y_d)$ given in \eqref{density2} and is given by \eqref{h1} and
{\small
\begin {align}
\nonumber
h(Y_d|X_s,X_r)=\dfrac{1}{M_{s}M_{r}}\sum_{i = 0}^{M_s-1}\sum_{j = 0}^{M_r-1}h(Y_d|X_s=x_{s,i},X_r=x_{r,j}),
\end {align}
}
\hspace{-0.255 cm} where $h(Y_d|X_s=x_{s,i},X_r=x_{r,j})$ can be calculated using the density function $P(Y_d=y_d|X_s=x_{s,i},X_r=x_{r,j})$ given in \eqref{density1} and it equals $h(z_d)$. The quantities $h(Y_d|X_s,X_r)$ and $I(X_s,X_r;Y_d)$ have been computed and are given by \eqref{h2} and \eqref{r1}.

\subsubsection{Calculation of $R_2$}

$I(X_s;Y_r|X_r)$ is given by,
{\small
\begin {align}
\nonumber
I(X_s,Y_r|X_r) = h(Y_r|X_r)-h(Y_r|X_s,X_r).
\end {align}
}
The quantity $h(Y_r|X_r=x_{r,i})$ can be evaluated using the density $P(Y_r|X_r=x_{r,j})$ given in \eqref{density4} and $h(Y_r|X_r)$ is given by,
{\small
\begin {align}
\nonumber
h(Y_r|X_r)=\dfrac{1}{M_{r}}\sum_{i = 0}^{M_r-1}h(Y_r|X_r=x_{r,i}).
\end {align}
} 
The quantities $h(Y_r|X_r,X_s)$ = $h(z_r)$, $h(Y_r|X_r)$ and $I(X_s;Y_r|X_r)$ have been computed and are given by \eqref{h3}, \eqref{h4} and \eqref{r2}.

\subsubsection{Calculation of $R_3$}

$I(X_s;Y_r,Y_d|X_r)$ is given by,
{\small
\begin {align}
\nonumber
&I(X_s;Y_r,Y_d|X_r) = h(Y_r,Y_d|X_r)-h(Y_r,Y_d|X_s,X_r),
\end{align}}
$\hspace{-0.255 cm}$ where $h(Y_r,Y_d|X_r)$ and $h(Y_r,Y_d|X_s,X_r)$ are given by,

{\small
\begin {align}
\nonumber
&h(Y_r,Y_d|X_r)=\dfrac{1}{M_{r}}\sum_{i = 0}^{M_r-1}h(Y_r,Y_d|X_r=x_{r,i});\\
\nonumber
&h(Y_r,Y_d|X_r,X_s)\\
\nonumber
&\hspace{1.5 cm}=\dfrac{1}{M_{s}M_{r}}\sum_{i = 0}^{M_r-1}\sum_{j = 0}^{M_s-1}h(Y_r,Y_d|X_r=x_{r,i},X_s=x_{s,j}).
\end {align}
}
The quantities $h(Y_r,Y_d|X_r=x_{r,i},X_s=x_{s,j})$ and $h(Y_r,Y_d|X_r=x_{r,i})$ can be evaluated using the density functions $P(Y_r,Y_d|X_r=x_{r,i},X_s=x_{s,j})$ and $P(Y_r,Y_d|X_r=x_{r,i})$ given in \eqref{density5} and \eqref{density6} (in the next page).

The quantities $h(Y_r,Y_d|X_r)$, $h(Y_r,Y_d|X_r,X_s)$ and $I(X_s;Y_r,Y_d|X_r)$ have been computed and are given by  \eqref{h5}, \eqref{h6} and \eqref{r3} (in the next page).

\newpage

\section{Calculation of bounds for the HD relay channel with finite input constellations}
\label{app-sec2}

\begin{figure*}
\scriptsize
\begin{align}
\label{density7}
&P(Y_r=y_r|X_s^1=x_{{s_1},i}) = \frac{1}{\pi} \exp\left(-\vert y_r-c_{rs}x_{s_1,i}\vert ^2\right)\\
\label{density8}
&P(Y_r=y_r) = \dfrac{1}{M_{s_1}}\sum_{i = 0}^{M_{s_1}-1}P(Y_r =y_r|X_s^1=x_{s_1,i})\\
\label{density9}
&P(Y_d^2=y_{d_2}|X_s^2=x_{s_2,i},X_r=x_{r,j}) = \frac{1}{\pi} \exp\left(-\vert y_{d,2}-c_{ds}x_{s_2,i}-c_{dr}x_{r,j}\vert ^2\right)\\
\label{density10}
&P(Y_d^2=y_{d,2}|X_r=x_{r,j}) = \dfrac{1}{M_{s,2}}\sum_{i = 0}^{M_{s_2}-1}P(Y_d^2 =y_{d,2}|X_{s,2}=x_{s_2,i},X_r=x_{r,j})\\
\label{density11}
&P(Y_d^2=y_{d,2}) = \dfrac{1}{M_{s,2}M_r}\sum_{i = 0}^{M_s-1}\sum_{j = 0}^{M_r-1}P(Y_d^2 =y_{d,2}|X_{s,2}=x_{s_2,i},X_r=x_{r,j})\\
\label{density12}
&P(Y_d^1=y_{d_1}|X_s^1=x_{{s_1},i}) = \frac{1}{\pi} \exp\left(-\vert y_{d,1}-c_{ds}x_{s_1,i}\vert ^2\right)\\
\label{density13}
&P(Y_d^1=y_{d_1}) = \dfrac{1}{M_{s_1}}\sum_{i = 0}^{M_{s_1}-1}P(Y_d^1=y_{d_1}|X_s^1=x_{s_1,i})\\
\label{density14}
&P(Y_r=y_r,Y_d^1=y_{d_1}|X_s^1=x_{s_1,i}) = \frac{1}{\pi^2} \exp\left(-\vert y_r-c_{rs}x_{s_1,i}\vert ^2-\vert y_{d_1}-c_{ds}x_{s_1,i}\vert ^2\right)\\
\label{density15}
&P(Y_r=y_r,Y_d^1=y_{d_1}) = \dfrac{1}{M_{s_1}}\sum_{i = 0}^{M_{s_1}-1}P(Y_r =y_r,Y_d^1=y_{d_1}|X_s^1=x_{s_1,i})\\
\hline
\label{r4}
&R_4~\triangleq~I(X_s^1;Y_r)]=log(M_{s_1})-\frac{1}{M_{s_1}}\sum_{i_1=0}^{M_{s_1}-1}\textbf{E}_{z_r}\left[log\left(\dfrac{\sum_{i = 0}^{M_{s_1}-1}exp\left(-\vert z_r-c_{rs}x_{s_1,i}+c_{rs}x_{s_1,i_1}\vert ^2\right)}{exp(-\vert z_r \vert^2)}\right)\right]\\
\label{r5}
&R_5~\triangleq~I(X_s^2;Y_d^2|X_r)]=log(M_{s_2})-\frac{1}{M_{s_2}}\sum_{i_1=0}^{M_{s_2}-1}\textbf{E}_{z_d^2}\left[log\left(\dfrac{\sum_{i = 0}^{M_{s_2}-1}exp\left(-\vert z_d^2-c_{ds}x_{s_2,i}+c_{ds}x_{s_2,i_1}\vert ^2\right)}{exp(-\vert z_d^2 \vert^2)}\right)\right]\\
\label{r6}
&R_6~\triangleq~I(X_s^1;Y_d^1)]=log(M_{s_1})-\frac{1}{M_{s_1}}\sum_{i_1=0}^{M_{s_1}-1}\textbf{E}_{z_d^1}\left[log\left(\dfrac{\sum_{i = 0}^{M_{s_1}-1}exp\left(-\vert z_d^1-c_{ds}x_{s_1,i}+c_{ds}x_{s_1,i_1}\vert ^2\right)}{exp(-\vert z_d^1 \vert^2)}\right)\right]\\
\nonumber
\label{r7}
&R_7~\triangleq~I(X_s^2,X_r;Y_d^2)=log(M_{s_2}M_r)\\  
& \hspace{3 cm} -\dfrac{1}{M_{s_2}M_r}\sum_{i_1 =0}^{M_{s_2}-1}\sum_{j_1 = 0}^{M_r-1}\textbf{E}_{z_d^2}\left[log\left(\dfrac{\sum_{i = 0}^{M_{s_2}-1}\sum_{j =0}^{M_r-1}exp\left(-\vert z_d^2-c_{ds}x_{s_2,i}-c_{dr}x_{r,j}+c_{ds}x_{s_2,i_1}+c_{dr}x_{r,j_1}\vert ^2\right)}{exp(-\vert z_d^2 \vert^2)}\right)\right]\\
\label{r8}
&R_8~\triangleq~I(X_s^1;Y_r,Y_d^1)]=log(M_{s_1})-\frac{1}{M_{s_1}}\sum_{i_1=0}^{M_{s_1}-1}\textbf{E}_{z_r ,z_d^1}\left[log\left(\dfrac{\sum_{i = 0}^{M_{s_1}-1}exp\left(-\vert z_r-c_{rs}x_{s_1,i}+c_{rs}x_{s_1,i_1}\vert ^2-\vert z_d^1-c_{ds}x_{s_1,i}+c_{ds}x_{s_1,i_1}\vert ^2\right)}{exp(-\vert z_r \vert^2-\vert z_d^1 \vert^2)}\right)\right]\\
\hline
\nonumber
\end{align}
\end{figure*}
The probability density functions used in the calculation of $R_4$-$R_8$ are given in \eqref{density7}-\eqref{density15} (in the next to the next page).
Following a procedure similar to the FD relay, using \eqref{density7}-\eqref{density15}, $R_4$-$R_8$ can be evaluated and are given by \eqref{r4}-\eqref{r8} (in the next to the next page).

\end{document}